\documentclass[useAMS,usenatbib]{mn2e}
\usepackage{natbib}
\usepackage{multirow}
\usepackage{graphicx}
\usepackage{txfonts}
\usepackage{lineno}

\title[Probing the VHE $\gamma$-ray spectral curvature in PG 1553+113]{Probing the very-high-energy $\gamma$-ray spectral curvature in the blazar PG 1553+113 with the MAGIC telescopes}

\author[J.~Aleksi\'c~et.~al.]{
J.~Aleksi\'c$^{1}$,
S.~Ansoldi$^{2}$,
L.~A.~Antonelli$^{3}$,
P.~Antoranz$^{4}$,
A.~Babic$^{5}$,
P.~Bangale$^{6}$, \newauthor 
J.~A.~Barrio$^{7}$,
J.~Becerra Gonz\'alez$^{8,25,}$
\thanks{Corresponding authors: 
J. Becerra Gonz\'alez, email: josefa.becerra@nasa.gov, 
P. Da Vela, email: davela@pi.infn.it,
E. Prandini, email: elisa.prandini@unige.ch,
F. D'Ammando, email: dammando@ira.inaf.it},
W.~Bednarek$^{9}$,
E.~Bernardini$^{10}$,
B.~Biasuzzi$^{2}$,\newauthor
% ...
 (for the {\it Fermi}-LAT Collaboration),
A.~L\"ahteenm\"aki$^{32, 33}$,
M.~Tornikoski$^{32}$, 
T.~Hovatta$^{32,34}$, \newauthor 
A.C.S.~Readhead$^{32}$,
W.~Max-Moerbeck$^{34}$,
J.L.~Richards$^{35}$ \newauthor 
(Affiliations can be found after the references)
}

\begin{document}
\label{firstpage}
\date{Accepted -  Received -; in original form -}

\pagerange{\pageref{firstpage}--\pageref{lastpage}} \pubyear{2014}

\maketitle

\begin{abstract}

PG 1553+113 is a very-high-energy  (VHE, $E>100\,\mathrm{GeV}$) $\gamma$-ray
emitter classified as a BL Lac object. Its redshift is constrained by
intergalactic absorption lines in the range $0.4<z<0.58$. The MAGIC telescopes
have monitored the source's activity since 2005. In early 2012, PG~1553+113
was found in a high-state, and later, in April of the same year, the source
reached its highest VHE flux state detected so far. Simultaneous observations carried out in X-rays during 2012 April show similar flaring behaviour. In contrast, the $\gamma$-ray flux at $E<100\,\mathrm{GeV}$ observed by {\it Fermi}-LAT is compatible with steady emission.
In this paper, a detailed study of the flaring state is presented. The VHE
spectrum shows clear curvature, being well fitted either by a power law with
an exponential cut-off or by a log-parabola. A simple power-law fit hypothesis
for the observed  shape of the PG\,1553+113 VHE $\gamma$-ray spectrum is
rejected with a high significance (fit probability P=2.6 $\times
10^{-6}$). The observed
curvature is compatible with the extragalactic background light (EBL) imprint
predicted by current generation EBL models assuming a redshift
$z\sim0.4$. New constraints on the redshift are derived from the VHE spectrum. These constraints are compatible with previous limits and suggest
that the source is most likely located around the optical lower limit,
$z=0.4$, based on the detection of Ly$\alpha$ absorption. Finally, we find that the synchrotron self-Compton (SSC) model gives a satisfactory description of the observed multi-wavelength spectral energy distribution during the flare.
\end{abstract}

\begin{keywords}
gamma rays: observations, blazar, BL Lac: AGNs: individual (PG~1553+113)
\end{keywords}


\section{Introduction}\label{introduction}

PG~1553+113 is a blazar found as part of the Palomar-Green Catalog of
UV-excess Stellar Objects \citep{green86}. Its J2000 coordinates are
R.A. 15h55m43.0s, Dec. +11d11m24.4s \citep{beasley}. It was classified as a BL
Lac object due to its featureless optical spectrum \citep{Miller1} and
significant optical variability \citep{Miller2}. As occurs in most BL Lac
objects, the featureless optical spectrum prevents a spectroscopic redshift
measurement. However, several limits have been provided based on indirect
measurements \citep[e.g.][]{Sbarufatti2005, Sbarufatti2006}. The most recent redshift lower limit was estimated assuming that the host galaxy can be used as a standard candle. For absolute R band magnitudes MR = -22.5 and MR = -22.9, \cite{shaw} obtains the limits $z>0.24$ and $z>0.31$, respectively. Previously, a more stringent redshift lower limit of $z>0.4$ was set by \cite{danforth10} based on the detection of intervening Ly$\alpha$ absorbers. This estimation will be used throughout the paper. \cite{danforth10} also set a redshift upper limit of $z<0.58$ based on the non-detection of any Ly$\beta$ absorbers at $z>0.4$.

The VHE $\gamma$-ray emission from PG~1553+113 was  discovered independently and almost simultaneously by H.E.S.S. \citep{Aharonian1} and MAGIC \citep{Albert1} in 2005.
The integral flux recorded by MAGIC at the time of the discovery was $F=(10.0\pm0.2_{stat})\times 10^{-11} \mathrm{cm}^{-2} \mathrm{s}^{-1}$ above 120\,GeV, and the differential energy spectrum was well described by a power law with a spectral index $\Gamma\sim 4$, compatible with the detection by H.E.S.S. 
The source has been monitored with the MAGIC telescopes since 2005. The results from the 2005-2009 observation campaigns can be found in \citet{Aleksic1}. Modest flux variability of a factor of $\sim$2.6 on a yearly time-scale has been detected at E$>$150~GeV, with an integral flux lying in the range from 1.4  to $3.7 \times 10^{-11} \mathrm{cm}^{-2} \mathrm{s}^{-1}$. 
The observed energy spectra were well fitted by power laws  with photon
indices in the range $\Gamma\sim3.6-4.3$ and compatible within uncertainties.

Extragalactic VHE $\gamma$-rays can be absorbed on the way to the Earth via electron-positron pair production when interacting with optical-UV background photons from the extragalactic background light \citep[EBL,][]{stecker_dejager,gould67}.  The EBL is mainly composed of diffuse optical light emitted by stars and partially reprocessed by dust in the IR, redshifted by the expansion of the Universe \citep{hauser}. 
The uncertainty on its spectral energy distribution (SED) and evolution through the history of the Universe still  ranges from 20\% to 50\% at wavelengths  0.4 and 40 microns, respectively. 
This uncertainty is mainly due to difficulties in direct measurements. 

During the past few years several different approaches have been developed to model the EBL \citep[e.g.][]{franceschini,KneiskeDole,finke,dominguez,gilmore2012, stecker2014} and despite the different techniques adopted the resulting EBL models show an overall agreement, differing only marginally.

The $\gamma$-ray absorption depends significantly on the energy of the VHE photon, the redshift-dependent SED of the EBL, and the distance to the source. The observed flux ($F_{obs}$) can be expressed as

\begin{equation}
F_{obs}(E)= F_{int}(E) \cdot e^{-\tau(E,z)} ,
\label{eq_EBL}
\end{equation}

where $F_{int}$ denotes the intrinsic flux emitted by the source and $\tau$ the EBL optical depth as a function of the energy and  redshift.

The EBL imprint on the VHE  $\gamma$-ray spectrum can be used to set upper limits on the redshift of the source. 
This is done by assuming a particular EBL model and a criterion on the intrinsic spectrum, such as a maximum hardness for the reconstructed intrinsic spectrum or the absence of a spectral break with a pile-up at VHE in the reconstructed spectrum.

Different authors have used this $\gamma$-ray attenuation technique  for
PG~1553+113, leading to the following limits: $z<0.74$ \citep{Aharonian2},
$z<0.42$ \citep{mazin}, $z<0.66$ \citep{Prandini}, $z\leq0.62$ \citep{Aliu}, $z=0.49\pm0.04$ \citep{abramowski15}, $z<0.53$ \citep{Biteau}. Limits on the EBL
absorption can be estimated independently from EBL models using the VHE
spectrum and the redshift of the source under the assumption that the emission
of the source can be properly described by a synchrotron self-Compton (SSC)
model \citep{Mankuzhiyil2010}. This method has previously been used on
PG~1553+113 to derive constraints on the $\gamma$-ray horizon
\citep{dominguez2013}. Alternatively, the EBL density relative to that of a given model can be evaluated through the joint analysis of the HE or VHE observations of many extragalactic sources, making relatively small assumptions on the shape of the intrinsic spectra \citep{abramowski13,ackermann12, Biteau}.

PG~1553+113 was detected in the high-energy (HE, $E>100$\,MeV) $\gamma$-ray band by the Large Area Telescope (LAT) on board the {\it Fermi Gamma-ray Space Telescope} \citep{Abdo2010}. The energy spectrum for the period 2008 August - 2010 August can be well fitted  by a power law with spectral index $\Gamma=1.67 \pm 0.02_{stat}$ and F($E>100\,\mathrm{MeV}$)=$(6.5 \pm 0.6_{stat}) \times 10^{-8}\mathrm{cm}^{-2} \mathrm{s}^{-1}$ and its variability index is 93.5 \citep{fermicat}. Since the variability index is $>$ 41.6 the source is variable on a monthly time-scale at $>$99\% confidence probability. No flaring activity has been claimed for PG~1553$+$113 in the HE band to date. Remarkably, a periodic factor-of-two flux enhancement on a monthly time scale over 6 years of Fermi operation has been recently found by \cite{ciprini}.

An extensive multi-wavelength (MWL) observation campaign on PG~1553+113 was carried out from 2012 February to June, focused on the characterization of its SED as well as the variability of the source emission at different frequencies. 
Observations from VHE $\gamma$-rays to radio were performed: VHE band by MAGIC, HE band by {\it Fermi}-LAT,  X-rays by {\it Swift}-XRT, UV-optical observations by {\it Swift}-UVOT, IR by REM and radio by Mets\"ahovi and OVRO. In this paper, the study of the flux variability in the VHE, HE and X-ray bands is presented. The study on the VHE spectrum is focused on the April flare state of the source. A detailed study on the long-term MWL campaign will be presented in a forthcoming paper.

The paper is organized as follows: Section~\ref{data} describes the MWL data analysis. In Section~\ref{results} the results are presented. The light curves from MAGIC, {\it Fermi}-LAT and {\it Swift}-XRT are shown in Section~\ref{Lightcurve} while a detailed analysis on the observed VHE energy spectrum is presented in Section~\ref{obs_spec}. The intrinsic VHE $\gamma$-ray spectrum together with a discussion on EBL imprint and redshift constraints can be found in Section~\ref{intrinsic}. The SED observed during the flaring state and the theoretical interpretation is described in Section~\ref{sed_section}. The conclusions can be found in Section~\ref{conclusions}.


\begin{figure*}[h!]
	\includegraphics[width=0.8\linewidth,clip]{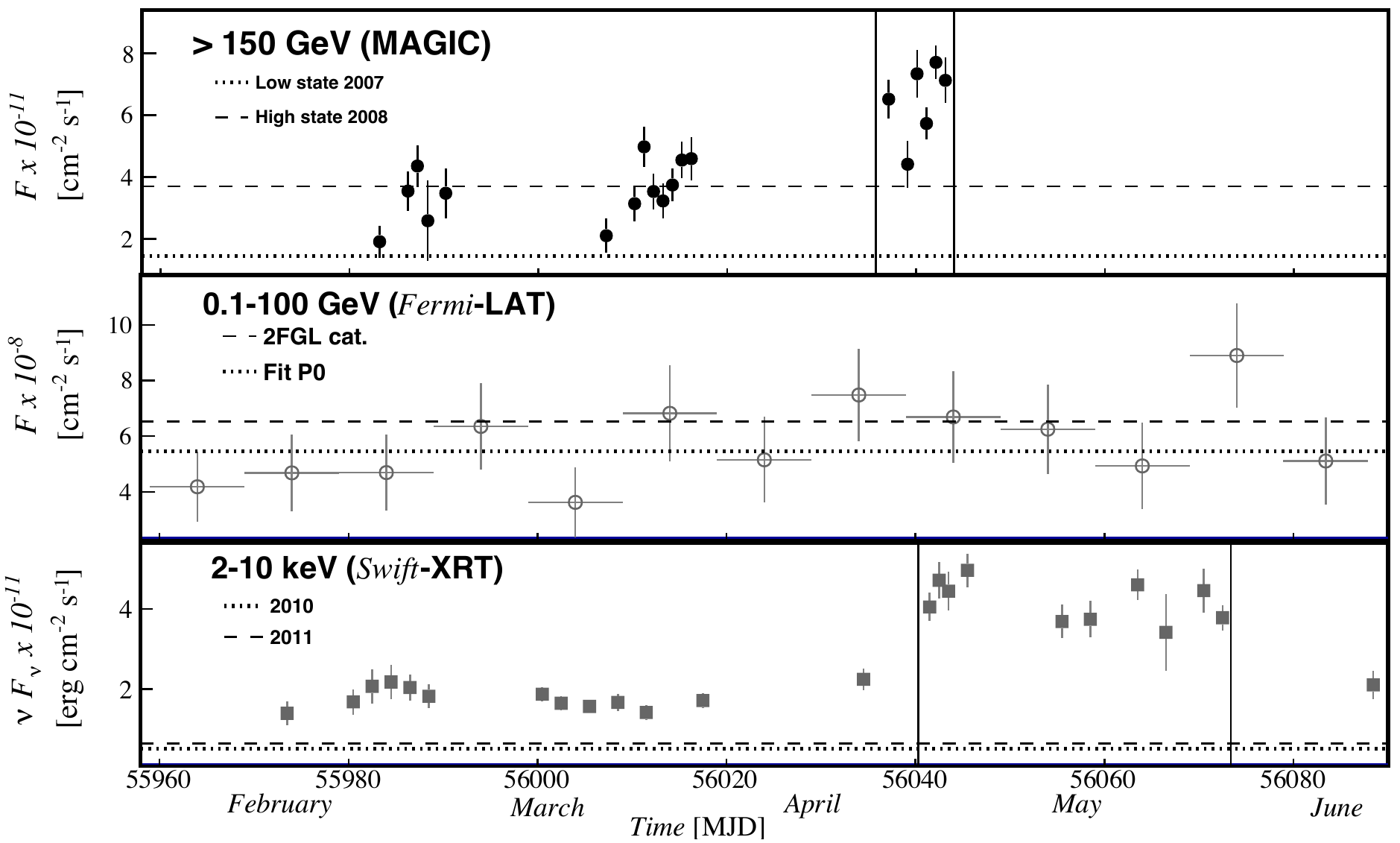}
	\caption{Light curve of PG~1553+113 during the 2012 observation campaign. Upper panel: nightly light curve in VHE $\gamma$-rays observed with the MAGIC telescopes for E$>$150\,GeV. For comparison the flux during the low state of the source in 2007 and the high state in 2008  measured by MAGIC  \citep{Aleksic1} are shown by the dotted and dashed lines, respectively. Middle panel: {\it Fermi}-LAT light curve computed with a 10-day binning for energies between 0.1 to 100\,GeV; the dashed gray lines show a zero order polynomial fit and the mean flux of the source from 2FGL catalog \citep{fermicat}. Lower panel: flux in the X-ray band in one night bins observed by {\it Swift}-XRT from 2 to 10 keV is represented in gray squares. For comparison previous measurements from 2010 and 2011 with {\it Swift}-XRT are plotted as dashed lines. The vertical lines denote the flare intervals observed in the VHE band and X-rays. No hints of flare are observed in the HE band.}
	\label{lc}
\end{figure*}

\section{Observations}\label{data}

\subsection{VHE $\gamma$-ray observations with MAGIC}

The VHE $\gamma$-ray observations were performed by the MAGIC telescopes. The MAGIC system consists of two 17\,m-diameter Imaging Atmospheric Cherenkov Telescopes (IACT) located on the Roque de los Muchachos, Canary Island of La Palma ($28^{\circ}46'\,$N,$17^{\circ}53'\,$W), at a height of 2200\,m above sea level. The system reaches a sensitivity of $(0.76\pm0.03)\%$ of the Crab Nebula flux for $E>290$\,GeV in 50~h of observations \citep{Aleksic2}.

PG~1553+113 was observed with the MAGIC telescopes from 2012 February 26 (MJD 55983) to April 26 (MJD 56043). The data sample after quality cuts consists of 18.3 h in the zenith angle range $17^\circ$ to $34^\circ$. The observations were performed in wobble mode \citep{fomin1994}, with the source located 0.4$^\circ$ from the centre of the field of view. The analysis of the data has been performed using the standard MAGIC analysis chain \citep{MARS1, MARS2}.
The energy threshold of the analysis is approximately 70 GeV.

The source was detected with a high statistical significance ($>70$ standard deviations, $\sigma$) during the time period 2012 February-April. The emission is compatible with a point-like source at the position of PG~1553+113. The mean $\gamma$-rate during the flare period is $4.35\pm0.04$\,$\gamma$/min for $E>70$\, GeV.

\subsection{HE $\gamma$-rays observations from {\it Fermi}-LAT}\label{fermi_section}

The {\it Fermi}-LAT  is a pair-conversion telescope operating from 20\,MeV to $>$ 300\,GeV. Further details about the {\it Fermi}-LAT can be found in
\citet{atwood09}. The LAT data reported in this paper were collected from 2012
February 2 (MJD 55959) to June 10 (MJD 56088). During this period, the
{\it Fermi} observatory operated almost entirely in survey mode. The analysis
was performed with the \texttt{ScienceTools} software package version
v9r32p5. The LAT data were extracted within a $10^{\circ}$ region of interest
centred at the location of PG\,1553$+$113. Only events belonging to the `Source' class were used. The time intervals when the rocking angle of the LAT was greater than 52$^{\circ}$ were rejected. In addition, a cut on the
zenith angle ($< 100^{\circ}$) was applied to reduce contamination from the Earth limb $\gamma$-rays, which are produced by cosmic rays interacting with the upper atmosphere.
The spectral analysis was performed with the instrument response functions
\texttt{P7REP\_SOURCE\_V15} using an unbinned maximum-likelihood method
implemented  in the Science tool \texttt{gtlike}. Isotropic (iso\_source\_v05.txt) and Galactic
diffuse emission (gll\_iem\_v05\_rev1.fit) components were used to model the
background\footnote{http://fermi.gsfc.nasa.gov/ssc/data/access/lat/Background\\Models.html} \citep{pass7}. The normalizations of both components in the background model
were allowed to vary freely during the spectral fitting.

We evaluated the significance of the $\gamma$-ray
signal from the source by means of the maximum-likelihood test
statistic TS = 2 (log$L_1$ - log$L_0$), where $L$ is the likelihood of the
data given the model with ($L_1$) or without ($L_0$) a point source at the
position of PG 1553$+$113 \citep[e.g.,][]{mattox96}. The model of the region of interest used in \texttt{gtlike} includes all point sources from the second {\it Fermi}-LAT catalogue
\citep[2FGL;][]{fermicat} as well as from a preliminary third {\it Fermi}-LAT
catalogue from 4 years of survey observations \citep{3FGL} that fall within $15^{\circ}$ radius around the source.
The
spectra of these sources were parametrized by power-law functions, except for
2FGL\,J1504.3.1$+$1023, 2FGL\,J1553.5$+$1255, and 2FGL\,J1608.5$+$1029, for which we used a log-parabola as in the 2FGL
catalogue. A first maximum-likelihood analysis was performed to remove from the
model sources having TS $<$ 10 and/or predicted number of counts based on
the fitted model $N_{pred} < 1$. A second maximum-likelihood analysis
was performed on the updated source model. In the fitting procedure, the normalization factors and the photon indices of the sources lying within 10$^{\circ}$ of PG\,1553$+$113 were left as free
parameters. For the sources located between 10$^{\circ}$ and 15$^{\circ}$, we
kept the normalization and the photon index fixed to the values from the 2FGL
catalogue. Integrating over the period from 2012 February 2 to June 10 (MJD 55959-56088) the fit with a power-law model in the 0.1--100 GeV energy range results in a TS = 908, with an integrated average flux of (5.7 $\pm$ $0.7_{stat}$) $\times$10$^{-8}$ cm$^{-2}$ s$^{-1}$ at the decorrelation energy of 2239 MeV and a photon index of $\Gamma$ = (1.59 $\pm$ $0.05_{stat}$) for PG\,1553$+$113. Using a log-parabola (LP), $dN/dE \propto$ $(E/E_{0})^{-\alpha-\beta \,\log(E/E_0)}$, the fit yielded for the same period a TS = 910, with an
average flux of (4.1 $\pm$ $0.8_{stat}$) $\times$10$^{-8}$ ph cm$^{-2}$ s$^{-1}$, a spectral slope of $\alpha$ = 1.49 $\pm$ $0.08_{stat}$ at the reference energy $E_0$
= 2239 MeV, and a curvature parameter $\beta$ = 0.08 $\pm$ $0.04_{stat}$. We used a likelihood ratio test (LRT) to check a PL model (null hypothesis) against a LP model (alternative hypothesis). These values may be compared, following \citet{fermicat}, by defining the curvature test statistic
TS$_{\rm curve}$=(TS$_{\rm LP}$ - TS$_{\rm PL}$). The LRT results in a TS$_{\rm curve}$ = 2, corresponding to a $\sim$1.4\,$\sigma$ difference. We, therefore, conclude, that no significant curvature was observed in the LAT spectrum of PG 1553$+$113 during 2012 February-June due to the poor photon statistics. Similar results were obtained when considering only photons with E$>$1 GeV. Above 10 GeV the analysis is strongly affected by the lack of statistics. No variability during the period was neither found. However, the Fermi-LAT spectrum of the source shows curvature when considering a longer integration time interval and we accumulate more photons at the highest Fermi-LAT energies. In the 3FGL catalog \citep{3FGL}, the spectrum of PG1553+113 is described by a log-parabola.

The $\gamma$-ray light curve using 10-day time bins and a PL model is reported in the middle
panel of Fig.~\ref{lc}. For each time bin, the spectral shape of
PG\,1553$+$113 and of all the sources within 10$^{\circ}$ of it were fixed
to the value obtained over the whole period.

The systematic uncertainty in the flux is dominated by the systematic uncertainty in the effective area \citep{pass7}. The systematic uncertainty on the effective area amounts to 10\% at 100\,MeV,
decreasing linearly with the logarithm of energy to 5\% between 316\,MeV and 10\,GeV, and increasing linearly with the logarithm of energy up to 10\% at 100\,GeV\footnote{http://fermi.gsfc.nasa.gov/ssc/data/analysis/LAT\_caveats.html}.

\subsection{X-rays and Optical-UV observations from {\it Swift}}\label{swift}

\begin{table*}
\caption{Log and fitting results of {\em Swift}-XRT observations of
  PG\,1553$+$113 using a log-parabola model with a HI column density fixed to
  the Galactic value in the direction of the source. Fluxes are unabsorbed.}
\begin{center}
\begin{tabular}{ccccccc}
\hline 
\multicolumn{1}{c}{Date} &
\multicolumn{1}{c}{Date} &
\multicolumn{1}{c}{Net exposure time} &
\multicolumn{1}{c}{$\alpha$} &
\multicolumn{1}{c}{$\beta$} &
\multicolumn{1}{c}{Flux 2.0--10 keV} &
\multicolumn{1}{c}{$\chi^2_{\rm red}$ (d.o.f.)} \\
\multicolumn{1}{c}{(MJD)} &
\multicolumn{1}{c}{(UT)} &
\multicolumn{1}{c}{(s)} &
\multicolumn{1}{c}{} &
\multicolumn{1}{c}{} &
\multicolumn{1}{c}{(10$^{-12}$ erg cm$^{-2}$ s$^{-1}$)} &
\multicolumn{1}{c}{}\\
\hline
55973 & 2012-02-16 & 2145 & 2.13 $\pm$ 0.11 & 0.31 $\pm$ 0.19 & $1.39 \pm 0.31$ & 1.018 (36) \\
55980 & 2012-02-23 & 2015 & 1.96 $\pm$ 0.10 & 0.27 $\pm$ 0.15 & $1.68 \pm 0.32$ & 0.902 (47) \\
55982 & 2012-02-25 & 1948 & 2.40 $\pm$ 0.10 & 0.19 $\pm$ 0.17 & $2.07 \pm 0.43$ & 0.867 (39) \\
55984 & 2012-02-27 & 2035 & 2.18 $\pm$ 0.10 & 0.06 $\pm$ 0.05 & $2.17 \pm 0.43$ & 1.129 (46) \\
55986 & 2012-02-29 & 1923 & 2.11 $\pm$ 0.10 & 0.26 $\pm$ 0.16 & $2.04 \pm 0.33$ & 1.098 (42) \\
55988 & 2012-03-02 & 2165 & 2.18 $\pm$ 0.09 & 0.45 $\pm$ 0.16 & $1.82 \pm 0.29$ & 0.999 (54) \\
56000 & 2012-03-14 & 1956 & 2.14 $\pm$ 0.06 & 0.40 $\pm$ 0.12 & $1.87 \pm 0.18$ & 0.930 (120) \\
56002 & 2012-03-16 & 2190 & 2.22 $\pm$ 0.05 & 0.37 $\pm$ 0.10 & $1.64 \pm 0.17$ & 1.198 (141) \\
56005 & 2012-03-19 & 2023 & 2.24 $\pm$ 0.05 & 0.49 $\pm$ 0.12 & $1.56 \pm 0.15$ & 1.019 (142) \\
56008 & 2012-03-22 & 1657 & 2.25 $\pm$ 0.06 & 0.44 $\pm$ 0.12 & $1.67 \pm 0.22$ & 1.208 (125) \\
56011 & 2012-03-25 & 1775 & 2.27 $\pm$ 0.06 & 0.41 $\pm$ 0.13 & $1.41 \pm 0.19$ & 1.053 (112) \\
56017 & 2012-03-31 & 2019 & 2.30 $\pm$ 0.05 & 0.29 $\pm$ 0.11 & $1.71 \pm 0.19$ & 1.082 (136) \\
56034 & 2012-04-17 &  912 & 2.20 $\pm$ 0.07 & 0.42 $\pm$ 0.14 & $2.24 \pm 0.27$ & 1.201 (87) \\
56041 & 2012-04-24 & 1062 & 2.12 $\pm$ 0.05 & 0.30 $\pm$ 0.11 & $4.05 \pm 0.35$ & 1.038 (136) \\
56042 & 2012-04-25 &  977 & 1.97 $\pm$ 0.06 & 0.43 $\pm$ 0.10 & $4.71 \pm 0.46$ & 1.099 (136) \\
56043 & 2012-04-26 &  996 & 2.17 $\pm$ 0.06 & 0.23 $\pm$ 0.12 & $4.44 \pm 0.49$ & 1.078 (102) \\ 
56045 & 2012-04-28 &  999 & 2.07 $\pm$ 0.05 & 0.42 $\pm$ 0.09 & $4.96 \pm 0.42$ & 0.964 (159) \\
56055 & 2012-05-08 &  529 & 2.10 $\pm$ 0.07 & 0.58 $\pm$ 0.14 & $3.68 \pm 0.42$ & 0.863 (83) \\ 
56058 & 2012-05-11 & 1184 & 2.15 $\pm$ 0.06 & 0.25 $\pm$ 0.12 & $3.74 \pm 0.45$ & 1.017 (108) \\ 
56063 & 2012-05-16 & 1098 & 2.11 $\pm$ 0.05 & 0.32 $\pm$ 0.10 & $4.60 \pm 0.39$ & 1.165 (155) \\ 
56070 & 2012-05-23 & 1256 & 2.09 $\pm$ 0.07 & 0.36 $\pm$ 0.13 & $4.46 \pm 0.54$ & 0.901  (96) \\ 
56072 & 2012-05-25 &  946 & 2.11 $\pm$ 0.06 & 0.37 $\pm$ 0.11 & $3.78 \pm 0.32$ & 1.078 (122) \\ 
56088 & 2012-06-10 &  955 & 2.19 $\pm$ 0.08 & 0.29 $\pm$ 0.16 & $2.07 \pm 0.35$ & 1.061  (70) \\ 
56091 & 2012-06-13 &  976 & 2.04 $\pm$ 0.07 & 0.52 $\pm$ 0.15 & $2.41 \pm 0.28$ & 0.965  (89) \\ 
56094 & 2012-06-16 &  602 & 2.14 $\pm$ 0.09 & 0.30 $\pm$ 0.19 & $2.45 \pm 0.45$ & 1.253  (59) \\ 
56095 & 2012-06-17 &  983 & 2.05 $\pm$ 0.07 & 0.42 $\pm$ 0.14 & $2.49 \pm 0.30$ & 0.916  (87) \\ 
56102 & 2012-06-24 &  556 & 2.00 $\pm$ 0.09 & 0.61 $\pm$ 0.19 & $2.66 \pm 0.42$ & 1.116  (60) \\ 
\hline
\end{tabular}
\end{center}
\label{XRT}
\end{table*}

{\it Swift} target of opportunity observations \citep{gehrels04}  of
PG~1553$+$113 were triggered by an increase of the flux emission observed in
the VHE band by the MAGIC telescopes \citep{atel1,atel2}. The  {\it Swift}
observations (XRT and UVOT) were performed in 2012 from February 16 (MJD 55973) to June 24 (MJD 56102). Previous
observations in 2010 and 2011 have also been used for comparison purposes. The
data taken with XRT on board {\it Swift} were processed with standard
procedures (\texttt{xrtpipeline v0.12.6}), filtering, and screening criteria
by using the \texttt{HEAsoft}\footnote{http://heasarc.nasa.gov/lheasoft/} package (v6.12). The data were collected both in photon counting (PC) and windowed timing (WT) mode, and XRT event grades 0--12 and 0--2 for the PC and WT events were selected, respectively \citep{burrows05}. Source events in WT mode were extracted from a circular region with a radius of 20 pixels (1 pixel $\sim$ 2.36''), while background events were extracted from a circular region with the same radius away from the source region.
Observations in PC mode showed an average count rate of  $>$ 0.5 counts s$^{-1}$, thus requiring pile-up correction. We extracted the source events
from an annular region with an inner radius of 5 pixels (estimated by means of the PSF fitting technique) and an outer radius of 30 pixels. We extracted
background events within an annular region centered on the source with radii 70 and 120 pixels. Ancillary response files were generated with xrtmkarf, and account for different extraction regions, vignetting and PSF corrections. We used the most recent spectral redistribution matrices in the calibration database maintained by HEASARC. We fit the spectrum with an absorbed log-parabola \citep[\texttt{logpar} in Xspec; e.g. ][]{massaro04} using the photoelectric absorption model \texttt{tbabs} \citep{wilms00}, with a neutral hydrogen column density fixed to its Galactic value \citep[3.65$\times$10$^{20}$ cm$^{-2}$, ][]{kalberla05}.

During the {\it Swift} pointings, the UVOT instrument observed PG
1553$+$113 in all its optical ($v$, $b$ and $u$) and UV ($w1$, $m2$ and
$w2$) photometric bands \citep{poole08,breeveld10}. We analysed the data
using the \texttt{uvotsource} task included in the \texttt{HEAsoft}
package. Source counts were extracted from a circular region of 5\arcsec\
radius centered on the source, while background counts were derived from a
circular region of 10\arcsec\ radius in the  source neighbourhood.
Conversion of magnitudes into de-reddened flux densities was obtained by
using the E(B-V) value of 0.046 from \citet{Schlafly}, the extinction
laws by \cite{cardelli98} and the magnitude-flux calibrations by
\citet{bessell98}.

\subsection{Infrared observations from REM}\label{REM}

PG 1553+113 was observed in the IR regime by the REM telescope from 2012 February 12 (MJD 55969) to July 30 (MJD 56138). The REM
\citep{Zerbi, Covino} is a robotic telescope located at the European Southern
Observatory (ESO) Cerro La Silla (Chile). It has a Ritchey-Chretien
configuration with a 60-cm f/2.2 primary and an overall f/8 focal ratio in a
fast moving alt-azimuth mount providing two stable Nasmyth focal stations. At
one of the two foci, the telescope simultaneously feeds, by means of a dichroic beamsplitter, two cameras: REMIR for the near-infrared band
\citep[NIR; ][]{Conconi} and REM Optical Slitless Spectrograph
\citep[ROSS,][]{Tosti} for the optical band. The cameras both have a field of view of 10 arcmin x 10 arcmin and imaging capabilities with the usual NIR (z, J, H and K) and Johnson-Cousins VRI filters. The REM software system \citep{Covino} is able to manage complex observational strategies in a fully autonomous way. All raw optical/NIR frames obtained with REM telescopes were reduced following standard procedures, i.e. dark frames obtained with the same exposure time are subtracted, sky flat-fields are applied and multiple dithered images are combined to derive sky frames. Multiple scientific frames are then combined to derive the final scientific images. Instrumental magnitudes were obtained via aperture photometry and absolute calibration has been performed by 2MASS objects in the field. The flux was corrected for Galactic reddening and extinction making use of \cite{Schlafly} extinction maps.

\subsection{Radio observations from Mets\"ahovi and OVRO}\label{radio}

PG~1553+113 was observed by the Mets\"ahovi 13.7-m radio telescope at 37 GHz during the MWL campaign from  2012 February 19 (MJD 55976) to March 24 (MJD 56010). The measurements were made with a 1 GHz-band dual beam receiver centered at 37 GHz. The observations are ON-ON observations, alternating the source and the sky in each feed horn. A detailed description of the observation and analysis methods can be found in \cite{Terasranta}. The detection limit (defined as S/N$\geq$4) of the telescope is of the order of 0.2 Jy under optimal weather conditions. Given the fact that the typical flux density of  PG~1553+113 is close to this limit, the source was significantly detected only on 2012 March 8 (MJD 55994) with a flux F=(0.20$\pm$$0.05_{stat}$) Jy.

The source is also monitored at 15\,GHz using the 40-m telescope of
the Owens Valley Radio Observatory (OVRO) as a part of a larger monitoring
program where a sample of $\sim$ 1700 sources are observed twice a
week \citep{richards11}. The telescope is equipped with
dual-beamed off-axis optics and a cooled receiver installed at the
prime focus. The two sky beams are Dicke switched using the off-source
beam as a reference, and the source is alternated between the two
beams in an ON-ON fashion to remove atmospheric and ground
contamination. Calibration is referenced to 3C~286 for which the flux
density of 3.44\,Jy at 15\,GHz is assumed \citep{baars77}. 
The systematic uncertainty is about 5\% in the flux density scale. Details on the observations, calibration and
analysis are given in \cite{richards11}.


\section{Results}\label{results}

In this section, a detailed analysis of the $\gamma$-ray and X-ray data is presented. 

\subsection{Flux variability}\label{Lightcurve}\label{lc_sec}

The light curves at VHE $\gamma$-rays, HE  $\gamma$-rays, and X-rays are shown in Fig.~\ref{lc}. For the VHE and X-rays bands, a nightly time-scale is used, while for the HE band we have used a 10-day binning. 
Clear variability is detected in both VHE and X-ray bands. The hypothesis of a constant flux can be rejected with high confidence level, P=$1.4\times10^{-21}$ ($\chi^2/ndf$=143.5/18) in VHE $\gamma$-rays and P=$1.7\times 10^{-50}$ ($\chi^2/ndf$=302.1/23) in X-rays. The HE flux is compatible with a constant flux of F=(5.5$\pm$$0.4_{stat}$)$\times 10^{-8}$ $\mathrm{cm}^{-2} \mathrm{s}^{-1}$ for energies 0.1--100\,GeV with a fit probability of P=0.6 ($\chi^2/ndf$=10.7/12). Note that the HE light curve is dominated by the emission at E$<$10\,GeV, accounting for 95\% of the photons.

In the VHE band, two states can be differentiated according to the source flux. In 2012 February-March the average source flux was at a level of F ($E>$150\,GeV)=(3.40$\pm$$0.15_{stat}$)$\times 10^{-11}$ $\mathrm{cm}^{-2} \mathrm{s}^{-1}$, corresponding to $\sim$11\% of the Crab Nebula flux measured by MAGIC \citep{crab08}. In 2012 April the source reached a flux above 150 GeV of (7.7$\pm$$0.5_{stat}$)$\times 10^{-11}$ $\mathrm{cm}^{-2} \mathrm{s}^{-1}$, approximately 24\% of the Crab Nebula flux.
Past MAGIC integral flux measurements above 150\,GeV lie in the range between 4\% (2007 observations) to 11\% (2008 observations) of the Crab Nebula flux as reported in \cite{Aleksic1}. Therefore we can conclude that in 2012 February-March the source was at a level comparable with a previously observed high state in 2008 (dashed line in the upper panel of Fig.~1). In 2012 April, instead, it reached the highest flux observed to date, about 6 times larger than the low state observed in 2007 and around twice that in February-March of the same year. According to the flux level, we divided the data into two samples: MJD 55983 to MJD 56016 (high state) and MJD 56037 to MJD 56043 (flare). The probability of a constant fit for both periods independently are low, P=$3.2\times 10^{-3}$ and P=$5.1\times 10^{-3}$, respectively. The shortest variability time scale observed is of the order of one day. No intra-night variability was detected. During the flare, the VHE flux approximately doubled with respect to the high state. The source was also observed by H.E.S.S. and VERITAS during this high state, and the results of these studies were reported during the publication of this manuscript \citep{abramowski15,Aliu}.

The X-ray flux observed in 2012 February-March in the 2-10 keV band  is compatible with a constant fit ($\chi^2/ndf$=12.7/12, P=0.4), with a mean flux $(1.71 \pm 0.06)\times10^{-11}$ $\mathrm{erg}$ $\mathrm{cm}^{-2} \mathrm{s}^{-1}$.  
In 2012 April-May the source was in a flare state compatible with a constant flux of $(4.20 \pm 0.14)\times10^{-11}$ $\mathrm{erg}$ $\mathrm{cm}^{-2} \mathrm{s}^{-1}$ ($\chi^2/ndf$=10.9/9, P=0.3). Later in 2012 June, the source flux  decreased to a level compatible with the flux measured during February-March (high state). 
For comparison, the flux measured in this band from previous observations during 2010 and 2011 was $(0.59\pm0.07_{stat})\times10^{-11}$ $\mathrm{erg}$ $\mathrm{cm}^{-2} \mathrm{s}^{-1}$, $(0.49\pm0.07_{stat})\times10^{-11}$ $\mathrm{erg}$ $\mathrm{cm}^{-2} \mathrm{s}^{-1}$ and $(0.64\pm0.13_{stat})\times10^{-11}$ $\mathrm{erg}$ $\mathrm{cm}^{-2} \mathrm{s}^{-1}$ measured during MJD 55198, MJD 55232 and MJD 55781, respectively. We can conclude that the X-ray flux doubled during the observation campaign. During the flare state the source reached a level of $\sim 7-10$ times the quiescent flux of the source measured during 2010 and 2011. 

\begin{figure}
	\includegraphics[width=1.03\linewidth,clip]{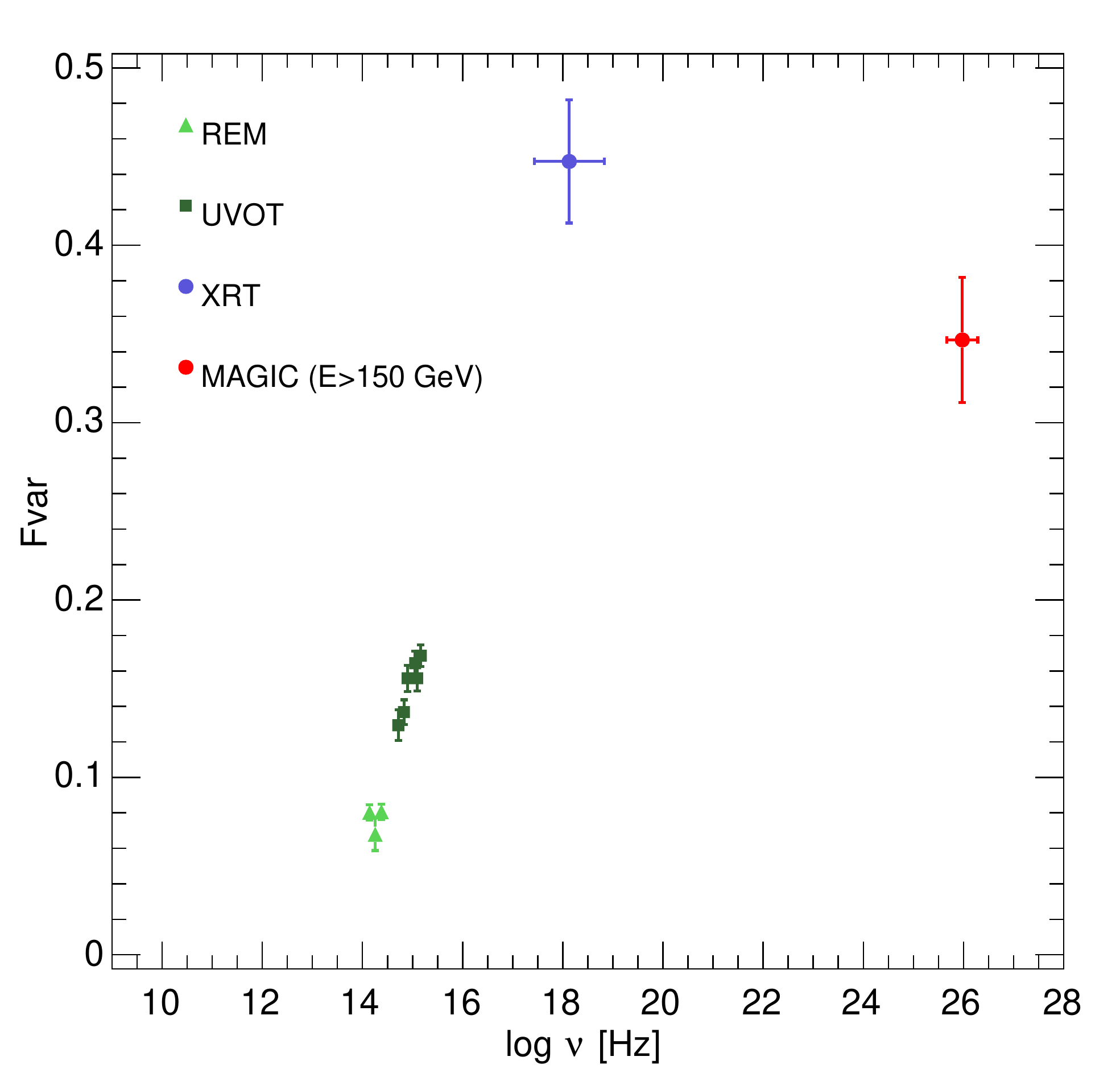}
	\caption{Fractional variability as a function of the frequency measured by different instruments.}
	\label{frac_var}
\end{figure}

Due to the lack of strictly simultaneous observations
it is difficult to perform an accurate comparison of  the VHE $\gamma$ and X-ray variability
properties. However, the flux evolution in both wavelengths suggests a correlation between the two bands. The SED can, in fact, be properly described in the framework of a one-zone Synchrotron Self Compton (SSC) model, pointing to a common origin of the emission in both energy bands as will be discussed in Section~\ref{sed}. Moreover, to quantify the intrinsic variability amplitude, the fractional variability was calculated at each measured frequency. The exception is the observations at 37\,GHz since the Mets\"ahovi observations resulted in one single detection. The fractional variability amplitude $F_{var}$ is defined as \citep{vaughan}: 

\begin{equation}
F_{var}=\sqrt{\frac{{S^2-<\sigma_{err}^2>}}{<F_{\gamma}>^2}}
\end{equation}

where $<F_{\gamma}>$ represents the average photon flux, S the standard deviation of the N flux measurements and $<\sigma_{err}^2>$ the mean squared error. $F_{var}$ is estimated for each frequency bin independently. The uncertainty on $F_{var}$ is calculated following the prescription from \cite{poutanen} as described in \cite{mrk501}:

\begin{equation}
\Delta F_{var}=\sqrt{F^2_{var}+err(\sigma^2_{NXS})}-F_{var}
\end{equation}

where $\sigma^2_{NXS}$ is given by equation 11 in \cite{vaughan}. The study was done only for the period in which we have the full MWL coverage (MJD 55983-56043). Daily fluxes have been used for all the frequencies except for the LAT, for which a 10-day bins have been used in order to detect the source in the individual bins. The fractional variability as a function of the frequency is shown in Fig.~\ref{frac_var} for those bands with positive excess variance ($S^2$ larger than $\sigma_{err}^2$). We obtained negative excess variance for the radio (OVRO) and HE gamma-ray ({\it Fermi}-LAT) bands, resulting in $F^2_{var}=-0.003$ and $F^2_{var}=-0.02$ respectively. Such negative excess variance is interpreted as absence of variability either because there was no variability, or because the instruments were not sensitive enough to detect it. Fig.~\ref{frac_var} shows clearly that the strongest intrinsic variability is observed in the X-rays ({\it Swift}/XRT) and the VHE (MAGIC) bands.

\subsection{X-ray to VHE spectral analysis}\label{obs_spec}

In X-rays PG 1553$+$113 showed a spectral curvature that can be well described with a log-parabolic shape \citep[see
  e.g.,][]{perlman05,tramacere07}. During the 2012 observation campaign the spectral index and curvature parameter varied in the range 2.0--2.4 and 0.2--0.6, respectively (Table~\ref{XRT}). Only for the observations performed
  on 2012 February 25 and 27 (MJD 55982 and 55984) no significant curvature seems to be present from the
  fit. No obvious connection was observed between the flux level and the
  curvature of the X-ray spectra.

As reported in Section~\ref{fermi_section}, no significant curvature was observed in the LAT
spectrum of PG 1553+113 during 2012 February-June.

In this paper, only the VHE $\gamma$-ray spectrum during the 2012 April flare (MJD 56037-56043) is presented, as mentioned in Section~\ref{introduction}. The VHE $\gamma$-ray spectra observed by MAGIC in 2012 February-March will be presented in a forthcoming paper.

\begin{figure}
   \includegraphics[width=1.\linewidth,clip]{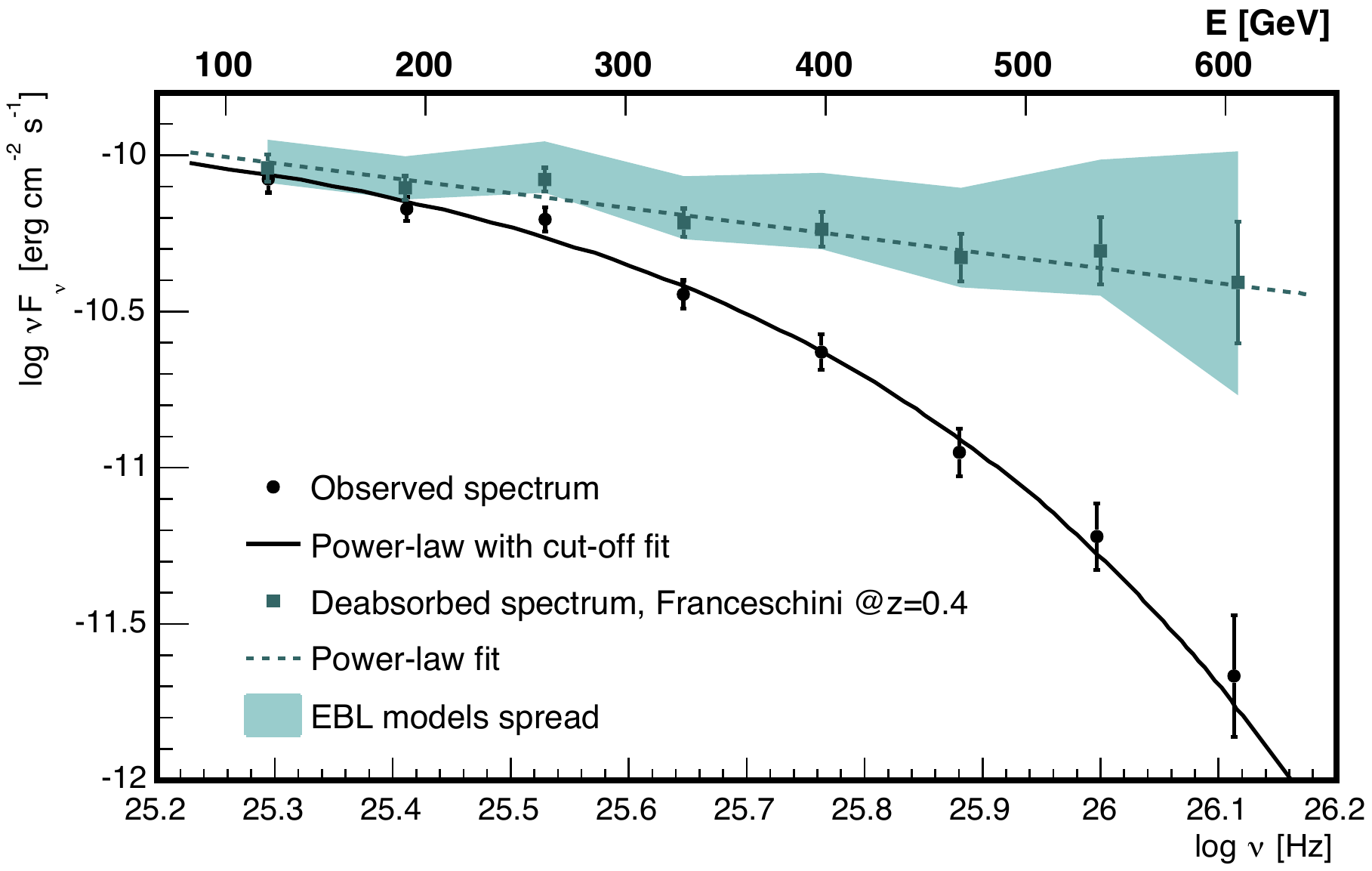}
   \caption{SED of PG~1553+113 as measured by MAGIC during the flare state of 2012 April. The observed SED is shown as black circles, and the black solid line represents the best fit to a power law with an exponential cut-off. The absorption-corrected spectrum assuming $z=0.4$ and using the EBL model by \citet{franceschini} is shown by the green squares; the dashed green line is the best-fitting power law. The green shaded area accounts for the uncertainties derived by the use of different EBL models.}
\label{fig:spectrum}
\end{figure}

The VHE spectrum during the flare is represented by black circles in
Fig.~\ref{fig:spectrum}. The differential VHE $\gamma$-ray spectral
  points can be found in Table~\ref{tab_spectrum}. They are corrected for instrumental effects by using the Schmelling unfolding algorithm \citep{Albert1}. 

The observed spectrum shows curvature, and a simple power-law fit can be discarded with a confidence level of 4.7 $\sigma$ ($P=2.6 \times 10^{-6}$, $\chi^2/ndf=36.1/6$). The differential spectrum can be well fit by a power law with an exponential cut-off with a probability of $P=0.7$ ($\chi^2/ndf=2.8/5$) in the energy range from $\sim$70~GeV to 620~GeV:

\begin{center}
\begin{equation}
\frac{dF}{dE}=f_0 \cdot \left(\frac{E}{200\,\mathrm{GeV}}\right)^{-\Gamma}
\cdot e^{-{E/E_c}} ,
\label{observed_spectrum}
\end{equation}
\end{center}

with a normalization constant of $f_0=(3.2 \pm 1.4_{stat} \pm 0.7_{sys}) \times 10^{-9} \mathrm{cm}^{-2} \mathrm{s}^{-1} \mathrm{TeV}^{-1}$, a photon index of $\Gamma=(1.87 \pm 0.37_{stat} \pm 0.15_{sys}) $ and $E_c=(110 \pm 24_{stat} \pm 19_{sys}$)~GeV. A full description of the systematics uncertainties for the MAGIC data analysis can be found in \cite{Aleksic2}.

The VHE $\gamma$-ray differential energy flux can be also well described by a log-parabola:

\begin{equation}
\frac{dF}{dE}=f_0 \cdot \left(\frac{E}{200\, \mathrm{GeV}}\right)^{-a-b \cdot \log{\frac{E}{200\,\mathrm{GeV}}}},
\end{equation}

where the parameters are given by a flux normalization constant at 200~GeV of  $f_0=(5.12 \pm 0.27_{stat} \pm 1.18_{sys} ) \times 10^{-10} \mathrm{cm}^{-2} \mathrm{s}^{-1} \mathrm{TeV}^{-1}$, $a=(3.83 \pm 0.10_{stat})$ and $b=(2.09 \pm 0.41_{stat})$, the systematic uncertainty on the spectral index is estimated to be $\pm0.15$. The goodness of the fit is given by $\chi^2/ndf=1.8/5$ with a probability $P=0.9$. 
A LRT shows that for the observed VHE differential spectrum a power law with an exponential cut-off and a log-parabola models are preferred with respect to a simple power law with significances of $5.8\,\sigma$ and $5.9\,\sigma$, respectively.

\begin{table*}
\begin{center}
\begin{tabular}{|cccc|} 
\hline
Energy bin&Energy&Flux& Flux uncertainty\\

[GeV]&[GeV]&[$\mathrm{TeV}^{-1}\times \mathrm{cm}^{-2} \times \mathrm{s}^{-1}$]&[$\mathrm{TeV}^{-1}\times \mathrm{cm}^{-2} \times \mathrm{s}^{-1}$]\\

\hline
71.2--93.4 & 81.5 &  7.90$\times 10^{-9}$   &  0.83$\times 10^{-9}$\\
93.4--91.1 &106.8 & 3.69$\times 10^{-9}$   &   0.34$\times 10^{-9}$\\
91.1--160.5 &139.9 & 1.99$\times 10^{-9}$  &   0.18$\times 10^{-9}$\\
160.5--210.5 &183.2 & 6.69$\times 10^{-10}$ &  0.73$\times 10^{-10}$\\
210.5--275.9 &239.9 & 2.55$\times 10^{-10}$ &   0.35$\times 10^{-10}$\\
 275.9--361.8 &314.0 & 7.10$\times 10^{-11}$ &  1.36$\times 10^{-11}$\\
361.8--474.3 &410.6 & 2.23$\times 10^{-11}$ &   0.62$\times 10^{-11}$\\
474.3--621.9 &536.6 & 4.68$\times 10^{-12}$ &   2.65$\times 10^{-12}$\\
\hline
\end{tabular}
 \caption{VHE differential energy spectra observed during the 2012 flare. First column represents the energy interval, the second the energy centre of each bin, the second the measured flux after unfolding and the last column is the flux uncertainty.}
 \label{tab_spectrum}
  \end{center}
\end{table*} 

\section{The intrinsic VHE $\gamma$-ray spectrum and the role of the
  EBL}\label{intrinsic}

\subsection{Origin of the curvature}

The VHE $\gamma$-ray spectrum is attenuated by the EBL, as described by Eq.~\ref{eq_EBL}. The optical depth ($\tau$) depends on the redshift of the VHE emitter and the energy of the $\gamma$-ray. In order to reconstruct the intrinsic spectrum emitted by a blazar, the redshift and the assumption of an EBL model is required. In the case of PG~1553+113, the uncertainty on the redshift prevents a precise estimation of the intrinsic spectrum. 
We adopt the optical lower limit from \cite{danforth10}, $z=0.4$, to study the EBL absorption effect in the observed spectrum, represented in Fig.~\ref{fig:spectrum}. 

The curvature measured in the observed VHE spectrum can have different
origins: intrinsic electron spectrum curvature, intrinsic
self-absorption, Klein-Nishina suppression and/or EBL absorption. The first hypothesis regarding the
energy distribution is not likely in the framework of the SED modeling as discussed in Section ~\ref{sed_section}. The assumption of the robust lower limit given by \cite{danforth10} allows us to test the possible contribution of intrinsic effect and EBL attenuation. 

Two possible scenarios can be envisioned considering the possible intrinsic absorption due to pair production within the source. If the $\gamma$-ray emission is produced within the broad line region (BLR) populated with optical-UV photons, a softening of the spectrum around tens of GeV would be expected \citep[e.g., ][]{Reimer2007,tavecchio2009,liu2006}. This is typically the case for flat spectrum radio quasars (FSRQ) showing strong optical emission lines, although usually weak for BL Lacs. In the far dissipation scenario \citep[e.g.,][]{sikora2008}, where the emission of $\gamma$-rays is assumed to be outside of BLR, the seed photons would come from the IR torus producing a softening in the spectrum at energies typically higher than 1\,TeV. None of these scenarios predict intrinsic absorption between 70 and 620\,GeV, especially from BL Lac objects with weak BLR emission.

The high flux of the source reached during the flare state allowed a high precision measurement of its spectrum. In addition, the spectrum extends to lower energies than previous measurements performed during lower flux states \citep{Aleksic1}. 
Despite the quality of the data and the high state of the source, no significant $\gamma$-ray emission was detected above 620~GeV, in agreement with previous measurements and with the $\gamma$-ray absorption expected by the state-of-the-art EBL models given the redshift limits. According to present generation of EBL models \citep{dominguez,KneiskeDole,franceschini,gilmore2012}, the observations during the flare reach an optical depth of $\tau\sim3$, which corresponds to $\sim$95\% photon absorption.

While the observed spectrum shows clear curvature, we find that  the spectrum corrected by the EBL effect assuming $z=0.4$ can be well described by a simple power law: 

\begin{center}
\begin{equation}
\frac{dF}{dE}=f_0 \cdot (\frac{E}{200\,\mathrm{GeV}})^{-\Gamma},
\label{deabsorbed_spectrum}
\end{equation}
\end{center}

whose parameters using the  \cite{franceschini} EBL model are given by a normalization flux at  200~GeV $f_0=(9.7 \pm 0.4_{stat} \pm 2.2_{sys}) \times 10^{-10} \mathrm{cm}^{-2} \mathrm{s}^{-1} \mathrm{TeV}^{-1}$ and a photon index of $\Gamma=(2.45 \pm 0.08_{stat} \pm 0.15_{sys})$. The probability of the fit is P=0.9 ($\chi^2/ndf=2.2/6$). The EBL-corrected spectrum is shown as green squares Fig.~\ref{fig:spectrum}, while the green shaded area represents the uncertainty when assuming different EBL models \citep{dominguez,KneiskeDole,franceschini,gilmore2012}.

\begin{figure}
  \includegraphics[width=0.95\linewidth,clip]{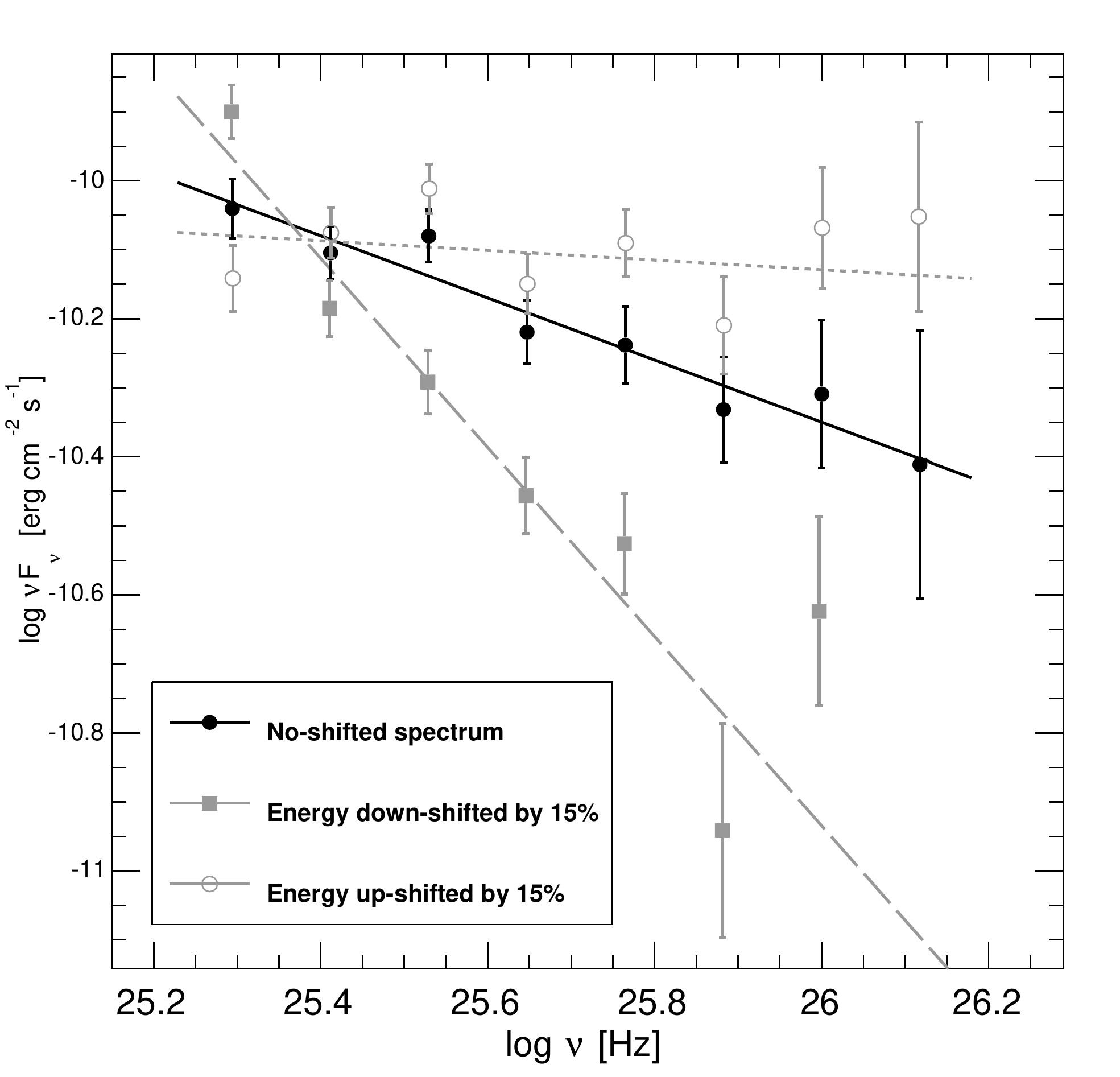}
  \caption{MAGIC spectral energy distribution EBL-corrected with ~\citet{franceschini} model by assuming $z=0.4$. The no-shifted spectrum is represented by the black circles. The solid grey squares show the spectrum considering a shift to lower energies by 15\%, and the grey open circles represent the spectrum accounting for a shift to higher energies by 15\%.}\label{shift_energy}
\end{figure}

We tested for a possible shift of 15\% in the energy scale due to the uncertainty in the energy measurement \citep{Aleksic2}. This was done by performing an event-wise shift in the data while leaving the Monte Carlo (MC) simulations (which are used to determine the energy of each event) unchanged. This simulates a data/MC mismatch, which could occur for numerous reasons including imperfect atmospheric conditions.  Both energy shifted spectra (see Fig.~\ref{shift_energy}), towards lower and higher, are compatible with a power-law fit with $\chi^2/ndf=6.7/5$ and $\chi^2/ndf= 8.1/6$, respectively. The shift to lower energies results in a steepening of the intrinsic spectrum (EBL-corrected according to \citet{franceschini} model assuming $z=0.4$) with a spectral index $\Gamma=3.37\pm0.12_{stat}$, while the shift to higher energies results in an intrinsic VHE $\gamma$-ray spectral index of $\Gamma=2.07\pm0.08_{stat}$.

The fact that the EBL-corrected VHE spectrum, assuming as a redshift the robust optical lower limit \citep{danforth10}, is compatible with a simple power law suggests that the curvature measured in the observed spectrum is very likely due to the interaction of the VHE photons with the EBL. 

\subsection{Redshift estimates} \label{sec:redshift}

\begin{figure}
   \includegraphics[width=0.95\linewidth,clip]{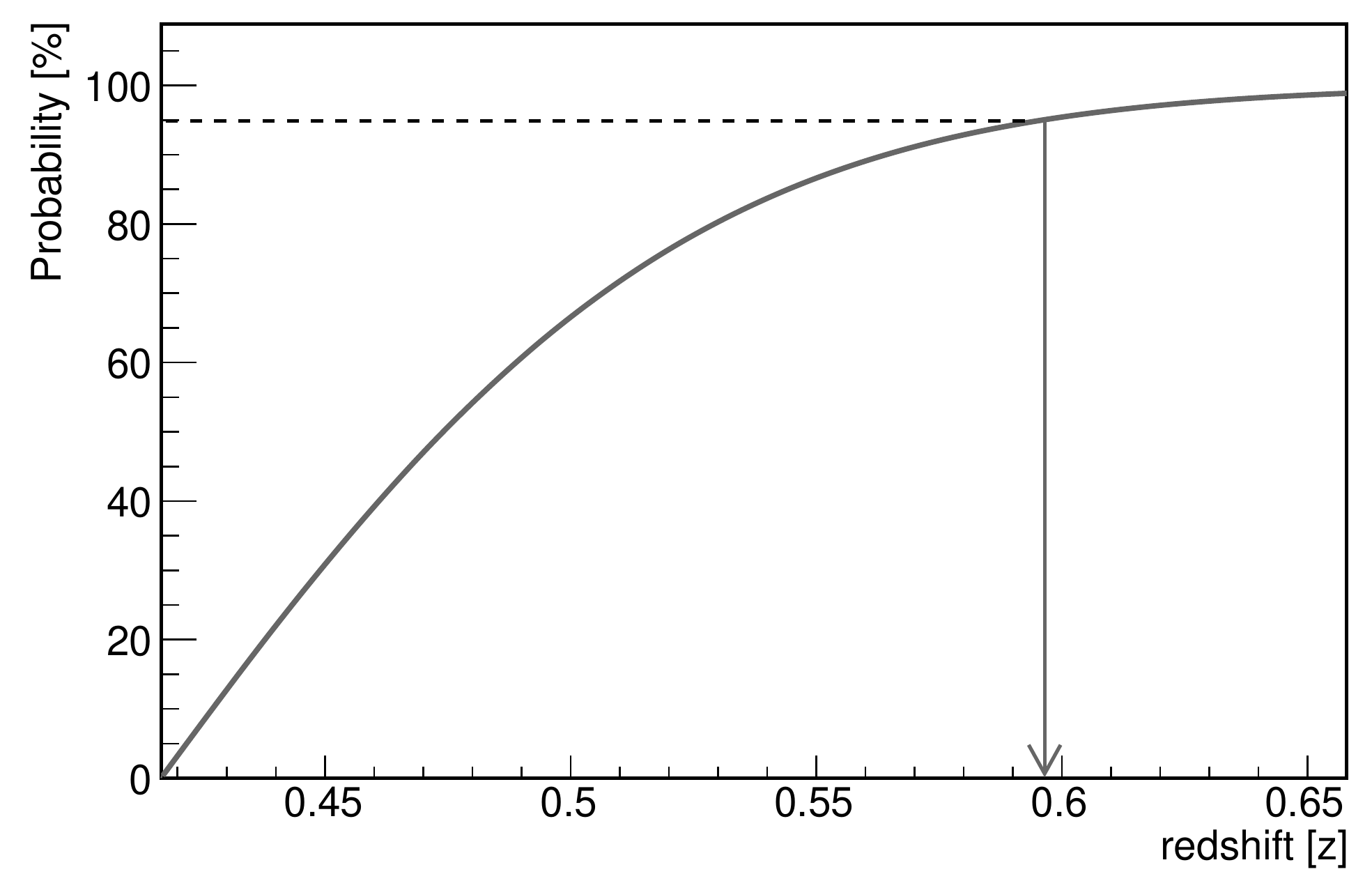}
   \caption{Constraint on the redshift of PG~1553+113 with the likelihood
     ratio test when comparing the hypothesis of a power-law fit and a
     positive curved power law (concave upward) fit for different distances of the source, using the EBL model from \citet{franceschini}.}
     	\label{fig:chi2ratio}
   \end{figure}

An upper limit on the redshift of PG~1553+113 was estimated by excluding the presence of a 
pile-up at high energies in the intrinsic VHE $\gamma$-ray spectrum. This approach is based on the underlying assumption that HE and VHE spectra connect smoothly and form the second peak of the SED. If the peak is located between 10 to 100\,GeV, as usually observed in TeV blazars, then a break between the HE and VHE spectra is expected, with VHE spectral slope softer than the HE slope. In the extreme case that the peak is located at higher energies, i.e. at some TeV, we expect that the HE and VHE spectra connect smoothly and exhibit the same spectral slope. A harder slope at VHE than at HE would imply the presence of an additional component in the SED, which is in general not expected, as discussed in ~\cite{ul}. Therefore, the redshift at which the two slopes equal after the correction for EBL absorption can be considered as an upper limit on the source distance under the assumption that there is no additional component. To find the upper limit, a LRT is performed. This test is used to evaluate the hypothesis of evidence of a break in the intrinsic spectrum, as proposed in \citet{mazin}. The hypothesis of a simple power-law fitting the EBL-corrected spectrum is compared with that of a curved power law, which can fit better the possible pile-up. 
For the PG~1553+113 data used in this work, the resulting probability is plotted in Fig.~\ref{fig:chi2ratio}. Above a redshift $\sim0.42$ a curved fit with positive curvature, which describes the pile-up, start to describe better the data than a simple power law. At redshift $z\geq$0.60 a curved fit with positive curvature is preferred to a simple power-law fit at the 95\% confidence level. Therefore the 
assumption that there is no spectral pile-up at high energies gives an upper 
limit on the source redshift of $z < 0.60$ within a 95\% confidence level.

\section{Spectral energy distribution}\label{sed_section}

\begin{figure}
  \includegraphics[width=0.95\linewidth,clip]{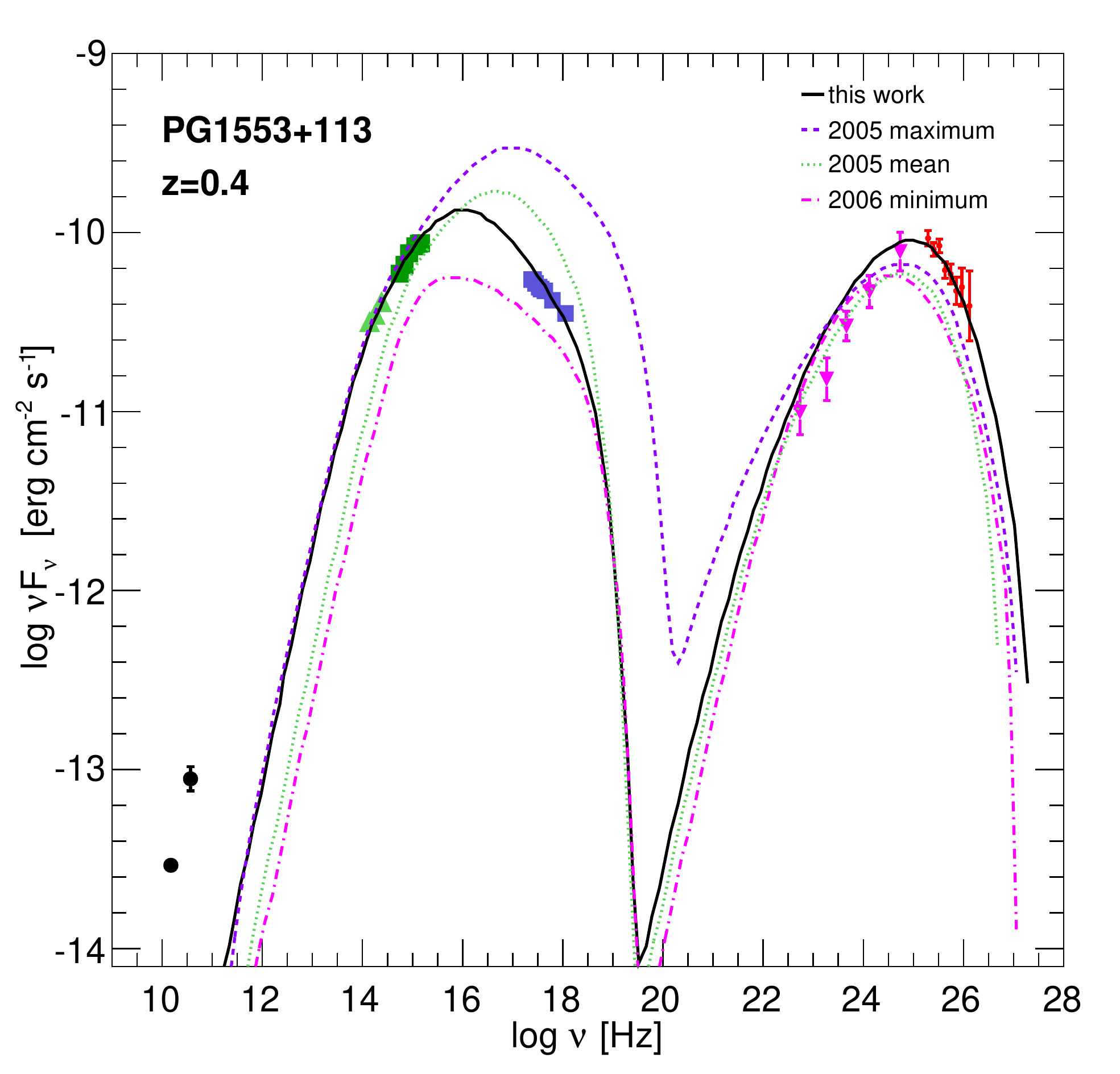}
  \caption{Spectral energy distribution of PG~1553+113 during the 2012 April flare state modeled with the one-zone SSC model of \citet{maraschi93}. From high to low energies: the EBL-corrected MAGIC spectra using \citet{franceschini} assuming z=0.4 (red dots, see text), the {\it Fermi}-LAT data from MJD 55959-56088 (pink triangles), {\it Swift}-XRT (purple squares) and {\it Swift}-UVOT (green squares) data from MJD 56045 (good representation of the X-ray and optical-UV state during the VHE flare), IR data from REM (green triangles) from MJD 56047, Mets\"ahovi (black square) from MJD 55994 (single detection) and mean radio observation in the period MJD 55959-56088 from OVRO (black circle). For comparison, the SSC models for previous source states \citep{Aleksic1} have been plotted in colored dashed lines.}\label{sed}
\end{figure}

Fig.~\ref{sed} shows the quasi-simultaneous SED observed during the flare state on 2012 April from $\gamma$-rays to radio. The VHE band is represented by the MAGIC observations during from MJD 56037-56043. The HE differential energy spectrum was derived using {\it Fermi}-LAT data, which covers the time interval from MJD~56030-56088. Because no variability was detected at HE by the LAT, we used a longer time interval to improve the statistics. The X-ray spectrum shown in Fig.~\ref{sed} represents the data collected by {\it Swift}-XRT on MJD~56045. The optical-UV data is the {\it Swift}-UVOT observation from the same day. 
The IR flux is estimated from REM telescope observations made on MJD~56046. At 37 GHz the single detection of the source by Mets\"ahovi on MJD 55994 is shown. The radio flux at 15 GHz measured by OVRO is compatible with a steady emission, therefore for the SED shown in Fig.~\ref{sed}, the mean flux from the period MJD 56037-56043 has been used.

The SED of PG~1553+113 data during the flare state has been modeled by using a one-zone SSC model \citep{maraschi93}. The emitting region is assumed to be spherical and populated by relativistic electrons. The electron spectrum is assumed to be a smoothed broken power law as a function of the energy (electron Lorentz factor) between $\gamma_{min}$ and $\gamma_{max}$ and break at $\gamma_b$:

\begin{equation}
N(\gamma)=K \gamma^{-n1} \left(1+\frac{\gamma}{\gamma_b}\right)^{n1-n2},
\label{electron_distribution}
\end{equation}

\noindent where K is the normalization factor, and n1 and n2 the spectral indices before and after the break. The region is filled with a tangled magnetic field and moves out of the jet with a given bulk Lorentz factor ($\Gamma$). The observable effect of bulk Lorentz factor depends on the viewing angle of the jet, which is taken into account in the Doppler factor ($\delta$) used for the SED modeling. 
According to the SSC model, the electrons emit synchrotron radiation due to their interaction with the magnetic field creating a low energy photon field, which can in turn interact with the same population of electrons via inverse Compton, producing the high energy emission. The synchrotron component considered is self-absorbed below $10^{11}$\ Hz and thus cannot reproduce the radio emission. This emission is likely from the superposition of multiple self-absorbed jet components \citep{konigl81}.

The parameters used for the modeling as well as those of SSC models reproducing previous observations \citep{Aleksic1} of the source in different states, for comparison purpose, can be found in Table~\ref{model}. During the strong flare in 2012 the magnetic field and the electron population normalization are significantly smaller than in previous states of the source, while the emitting region size is six times larger. However, as given by the causality relation $\mathrm{R}<(\mathrm{c}\cdot \mathrm{t} \cdot  \delta) /(1+z)$, the allowed flux variability time-scale is $\sim19$\,h (assuming z=0.4), which is compatible with the variability detected in the source as shown in Fig.~\ref{lc}. The inverse Compton (IC) energy peak moved to higher energies with respect to previous observations and more energetic particles were involved, requiring a larger $\gamma_{max}$ parameter for the modeling as shown in Table ~\ref{model}. This could point to different origins of the high states of the source. The derived luminosities from the SSC modeling are shown in Table~\ref{luminosities}. The electron and (cold) proton luminosities are higher than previous high states. For the luminosities calculation one proton per emitting electron was assumed. It is worth noting that the MWL data of the previous states of the source \citep{Aleksic1} used for comparison were not simultaneous. Moreover, due to the degeneracy of the model parameters the best model is not unique and other parameters could also reproduce the SED. Therefore, strong conclusions cannot be derived from the comparison with previous modeling of the source.

As shown in Fig.~\ref{sed}, the increasing part of both SED bumps shows less variability when compared with the decreasing part. This fact is also in agreement with the light curve discussion on Section~\ref{lc_sec}: while X-rays and VHE $\gamma$-rays show an increase of the flux in 2012, the emission in the HE band is compatible with a constant flux. The high variability found in X-rays and VHE $\gamma$-rays suggests that the flaring activity of this source is driven by the most energetic electrons. Moreover, as discussed previously, the SSC model gives a lower magnetic field with respect to previous models, which implies a longer synchrotron cooling time-scale. This is in agreement with the displacement of the synchrotron peak to higher frequencies, as well as with the higher variability in the high energy component of both peaks.

As shown in Fig.~\ref{sed}, the IC peak of the SED is close to the VHE
band. Therefore, curvature would be expected in the intrinsic VHE SED due to
the distribution of the relativistic electrons, within the one-zone SSC
framework (as mentioned in Sec.~\ref{intrinsic}). To test if our observations
are sensitive enough to detect the expected intrinsic SED curvature, we
simulate the MAGIC response assuming the intrinsic emission given by the best SSC modeling of the MWL data shown in Fig.~\ref{sed}. We simulate
intrinsic VHE SEDs assuming the same frequency binning and relative errors as
in the observed VHE spectral points (only statistical uncertainties have been
taken into account). The result of ten thousand realizations are shown in
Fig.~\ref{ssc_simulation}, and are represented by the gray shaded
area. Despite the simulated SEDs having, by construction, an evident
curvature, 99.2\% of the realizations are well described by a simple power law. To be conservative, we require a p-value of the individual fits $>$0.9973, which allows us to exclude spectral curvature in the simulated spectra at a 3$\sigma$ confidence level. The mean probability of a simple power-law fit
is P = 0.44$\pm$0.28 with a mean photon index of $2.38 \pm 0.10$. We therefore conclude that the sensitivity of our VHE measurements do not allow the detection of an intrinsic curvature in the SSC framework and the EBL model from \cite{franceschini}.

\begin{table*}
\begin{center}
\begin{tabular}{|c|ccccccccccccc|} 
\hline
Model&$\gamma _{\rm min}$ & $\gamma _{\rm b}$ & $\gamma _{\rm max}$ &
$n_1$ & $n_2$ &$B$ & $K$ &$R$ & $\delta $ \\
&$[10^3$] & [$ 10^4$] &[$ 10^5$]  &  & &[G] & [cm$^{-3}]$  & $[10^{16}$cm] &\\
\hline
This work&3.7&3.6&8.0&1.60&3.83&0.045&19.5&6.0&40\\
\hline
$\mathrm{Maximum}^{a}$ &1.0&3.0&5.2&2.00&3.75&0.800&$3.8 \times 10^{3}$&1.0&35\\

\hline
$\mathrm{Minimum}^{a}$ &5.0&1.3&4.1&2.00&3.55&0.200&$25.0 \times 10^{3}$&1.0&35\\

\hline
$\mathrm{Mean}^{a}$ &1.5&3.2&2.2&2.00&4.00&0.500&$5.4 \times 10^{3}$&1.0&35\\
\hline
\end{tabular}
\caption{ One-zone SSC model parameters of the SED fit during the flare state on 2012. The models marked as $^{a}$  correspond to previous activity states of the source \citep[see ][]{Aleksic1} and are shown for comparison. The following quantities are reported: the minimum, break, and maximum Lorentz factors and the low and high energy slope of the electron energy distribution, the magnetic field intensity, the electron density, the radius of the emitting region and the Doppler factor. The derived luminosities are shown in Table~\ref{luminosities}.}
\label{model}
\end{center}
\end{table*}

\begin{table*}
\begin{center}
\begin{tabular}{|c|cccc|} 
\hline 
Model&L$_{kin(e)}$& L$_{kin(p)}$& L$_B$ & log10($\nu_{\mathrm{syn}}$)\\
&[$10^{45}$ erg s$^{-1}$]&[$10^{44}$ erg s$^{-1}$]&[$10^{43}$ erg s$^{-1}$] &\\
\hline
This work&2.18&1.49&5.83&16.1\\
\hline
$\mathrm{Maximum}^{a}$ &0.52&0.6&39.2&17.0\\
\hline
$\mathrm{Minimum}^{a}$ &0.52&0.6&2.5&15.9\\
\hline
$\mathrm{Mean}^{a}$ &0.52&0.6&15.3& 16.7\\
\hline
\end{tabular}
\caption{ Luminosities derived from the one-zone SSC model of the SED during the flare state on 2012. The models marked as $^{a}$  correspond to previous activity states of the source \citep[see ][]{Aleksic1} and are shown for comparison. The following quantities are reported: the kinetic energy of the electrons, (cold) protons (assuming one proton per emitting electron), and magnetic field, and the frequency of the synchrotron peak. The model parameters are show in Table~\ref{model}.}
\label{luminosities}
\end{center}
\end{table*}

\begin{figure}\label{ssc_simulation}
  \includegraphics[width=0.95\linewidth,clip]{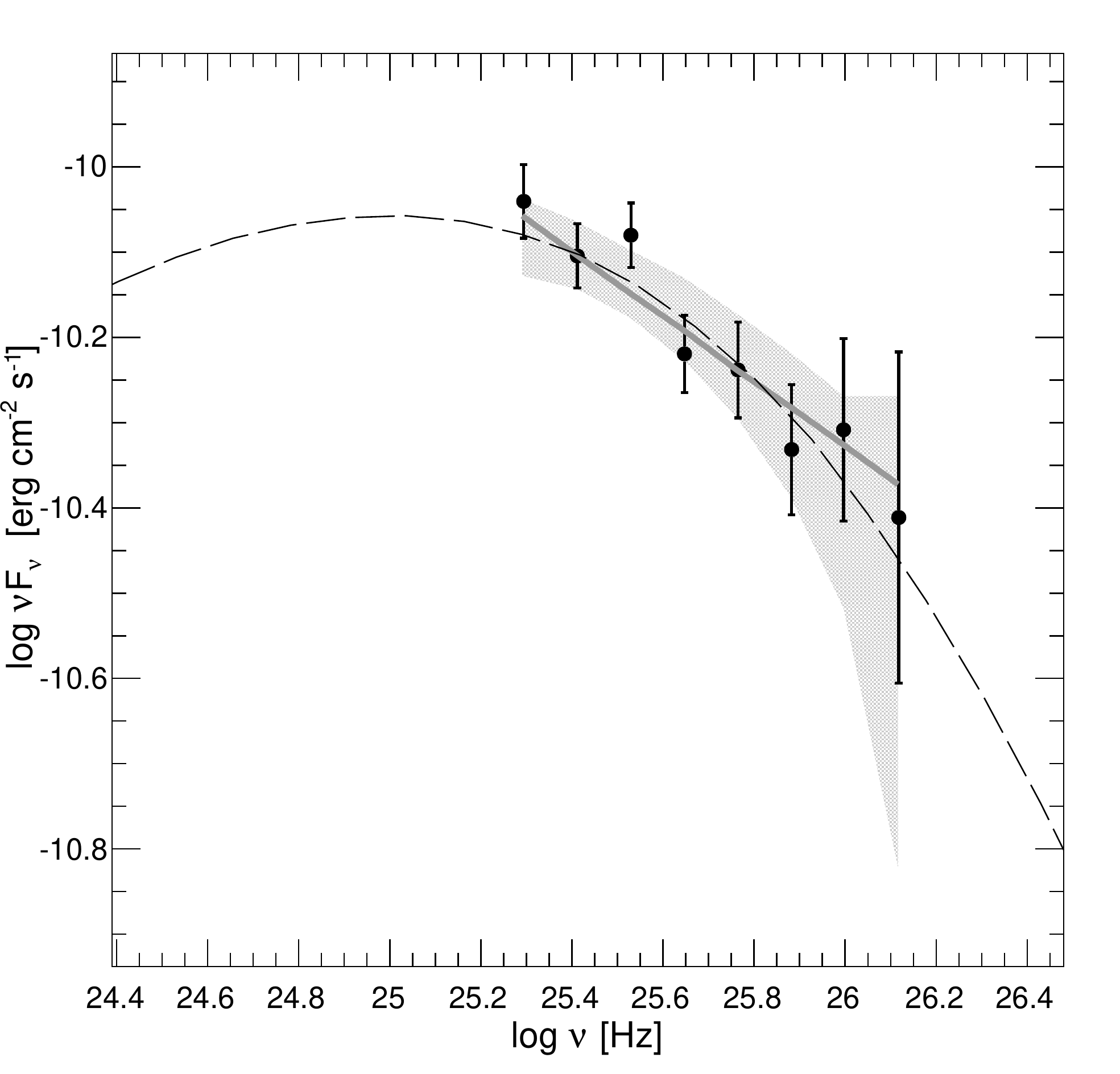}
  \caption{PG~1553+113 VHE SED. The gray shaded area represents the simulated MAGIC response assuming the intrinsic emission given by the best-fitting SSC model to the MWL data shown in Fig.~\ref{sed}. The gray solid line represents the mean power-law fit of the ten thousand realizations of the toy-MC. The black circles denote the PG\,1553+113 VHE spectrum EBL-corrected with ~\citet{franceschini} model assuming z=0.4. The dashed black line represent the best-fitting SSC model from Fig.~\ref{sed}.}\label{ssc_simulation}
 \end{figure}

\section{Conclusions}\label{conclusions}

In this paper we have presented the highest flux state ever detected from the blazar PG 1553+113 in VHE $\gamma$-rays.
The flare was detected at VHE by the MAGIC telescopes and monitored in HE
$\gamma$-rays by {\it Fermi}-LAT, in X-rays by {\it Swift}-XRT, in optical-UV
by {\it Swift}-UVOT, in infrared by REM, and in radio by Mets\"ahovi and
OVRO. While clear variability has been found in both the VHE and X-ray bands, the HE $\gamma$-ray flux is compatible with constant emission.

The observed VHE spectrum shows clear curvature and a simple power-law fit is rejected with a confidence level of 4.7$\sigma$. It is well fitted by a power law with an exponential cut-off or a logparabola. This curvature most likely originates from EBL absorption if the distance to the source is between the redshift limits measured by \citet{danforth10} ($0.4< z< 0.58$). If the real redshift of this source is higher than these limits, the effect would be a hardening of the spectrum or the (unexpected) presence of a pile-up in the intrinsic spectrum, which would denote that either the EBL models predict an overestimated EBL level, there is a second emission component at high energies or that more exotic physics needs to be invoked as axion-like particles (ALPs) \citep[e.g.,][]{DeAngelis, hooper, ALPs} or the effect of Lorentz invariance violation \citep[e.g.][]{jacob}.

A redshift upper limit of $z<0.60$ at 95\% C.L. has been derived using the $\chi^2$ 
ratio test \citep{mazin}. A quasi-simultaneous SED has been compiled for the flare episode in 2012 April. 
It can be well modeled by a one-zone SSC model. 
The comparison with previous flux states of the source reveals that the higher frequency part of each SED bump shows higher variability than the lower frequency part. This fact points to a scenario where the most energetic electrons play a leading role during the flare episodes of the source.

A detailed study of the MWL behaviour and evolution of the SED will 
be published in a forthcoming paper.


\section*{Acknowledgements}
We would like to thank the Instituto de Astrof\'{\i}sica de Canarias for the excellent working conditions at the Observatorio del Roque de los Muchachos in La Palma. The financial support of the German BMBF and MPG, the Italian INFN and INAF,  the Swiss National Fund SNF, the ERDF under the Spanish MINECO, and the Japanese JSPS and MEXT is gratefully acknowledged. This work was also supported by the Centro de Excelencia Severo Ochoa SEV-2012-0234, CPAN CSD2007-00042, and MultiDark CSD2009-00064 projects of the Spanish Consolider-Ingenio 2010 programme, by grant 268740 of the Academy of Finland, by the Croatian Science Foundation (HrZZ) Project 09/176 and the University of Rijeka Project 13.12.1.3.02, by the DFG Collaborative Research Centers SFB823/C4 and SFB876/C3, and by the Polish MNiSzW grant 745/N-HESS-MAGIC/2010/0.
Part of this work has been possible with the support of the Cluster of Excellence: "Connecting
Particles with the Cosmos", part of the Landesexzellenzinitiative Hamburg. The authors thank to S.~Buson, J.~Finke, J.~Perkins and A.~Neronov for their comments and contributions. JBG would like to thanks to M. Raue for  useful discussions.

The {\it Fermi} LAT Collaboration acknowledges generous ongoing
support from a number of agencies and institutes that have supported both
the development and the operation of the LAT as well as
scientific data analysis.  These include the National Aeronautics and
Space Administration and the Department of Energy in the United
States, the Commissariat \`a l'Energie Atomique and the Centre
National de la Recherche Scientifique / Institut National de Physique
Nucl\'eaire et de Physique des Particules in France, the Agenzia
Spaziale Italiana and the Istituto Nazionale di Fisica Nucleare in Italy,
the Ministry of Education, Culture, Sports, Science and
Technology (MEXT), High Energy Accelerator Research Organization (KEK) and
Japan Aerospace Exploration Agency (JAXA) in Japan, and the
K.~A.~Wallenberg Foundation, the Swedish Research Council and the
Swedish National Space Board in Sweden. Additional support for science
analysis during the operations phase is gratefully acknowledged from the
Istituto Nazionale di Astrofisica in Italy and the Centre National
d'\'Etudes Spatiales in France.

We thank the {\it Swift} team duty scientists and science planners for making these observations possible;

The OVRO 40-m monitoring program is supported in part by NASA grants
NNX08AW31G and NNX11A043G, and NSF grants AST-0808050 and AST-1109911.

The Mets\"ahovi team acknowledges support from the Academy of Finland to
our observing projects (numbers 212656, 210338, 121148, and others).

\vspace*{0.5cm}
\noindent
$^{1}$ {IFAE, Campus UAB, E-08193 Bellaterra, Spain} \\
$^{2}$ {Universit\`a di Udine, and INFN Trieste, I-33100 Udine, Italy} \\
$^{3}$ {INAF National Institute for Astrophysics, I-00136 Rome, Italy} \\
$^{4}$ {Universit\`a  di Siena, and INFN Pisa, I-53100 Siena, Italy} \\
$^{5}$ {Croatian MAGIC Consortium, Rudjer Boskovic Institute, University of Rijeka and University of Split, HR-10000 Zagreb, Croatia} \\
$^{6}$ {Max-Planck-Institut f\"ur Physik, D-80805 M\"unchen, Germany} \\
$^{7}$ {Universidad Complutense, E-28040 Madrid, Spain} \\
$^{8}$ {Inst. de Astrof\'isica de Canarias, E-38200 La Laguna, Tenerife, Spain} \\
$^{9}$ {University of \L\'od\'z, PL-90236 Lodz, Poland} \\
$^{10}$ {Deutsches Elektronen-Synchrotron (DESY), D-15738 Zeuthen, Germany} \\
$^{11}$ {ETH Zurich, CH-8093 Zurich, Switzerland} \\
$^{12}$ {Universit\"at W\"urzburg, D-97074 W\"urzburg, Germany} \\
$^{13}$ {Centro de Investigaciones Energ\'eticas, Medioambientales y Tecnol\'ogicas, E-28040 Madrid, Spain} \\
$^{14}$ {Institute of Space Sciences, E-08193 Barcelona, Spain} \\
$^{15}$ {Universit\`a di Padova and INFN, I-35131 Padova, Italy} \\
$^{16}$ {Technische Universit\"at Dortmund, D-44221 Dortmund, Germany} \\
$^{17}$ {Unitat de F\'isica de les Radiacions, Departament de F\'isica, and CERES-IEEC, Universitat Aut\`onoma de Barcelona, E-08193 Bellaterra, Spain} \\
$^{18}$ {Universitat de Barcelona, ICC, IEEC-UB, E-08028 Barcelona, Spain} \\
$^{19}$ {Japanese MAGIC Consortium, Division of Physics and Astronomy, Kyoto University, Japan} \\
$^{20}$ {Finnish MAGIC Consortium, Tuorla Observatory, University of Turku and Department of Physics, University of Oulu, Finland} \\
$^{21}$ {Inst. for Nucl. Research and Nucl. Energy, BG-1784 Sofia, Bulgaria} \\
$^{22}$ {Universit\`a di Pisa, and INFN Pisa, I-56126 Pisa, Italy} \\
$^{23}$ {ICREA and Institute of Space Sciences, E-08193 Barcelona, Spain} \\
$^{24}$ {Universit\`a dell'Insubria and INFN Milano Bicocca, Como, I-22100 Como, Italy} \\
$^{25}$ {now at: NASA Goddard Space Flight Center, Greenbelt, MD 20771, USA and Department of Physics and Department of Astronomy, University of Maryland, College Park, MD 20742, USA}\\
$^{26}$ {now at Ecole polytechnique f\'ed\'erale de Lausanne (EPFL), Lausanne, Switzerland}\\
$^{27}$ {now at Institut f\"ur Astro- und Teilchenphysik, Leopold-Franzens- Universit\"at Innsbruck, A-6020 Innsbruck, Austria}\\
$^{28}$ {now at Finnish Centre for Astronomy with ESO (FINCA), Turku, Finland}\\
$^{29}$ {also at INAF-Trieste}\\
$^{30}$ {also at ISDC - Science Data Center for Astrophysics, 1290, Versoix (Geneva)}\\
$^{31}$ {INAF-IRA, I-40129 Bologna, Italy}\\
$^{32}$ Aalto University Mets\"ahovi Radio Observatory, Mets\"ahovintie 114, 02540, Kylm\"al\"a, Finland\\
$^{33}$ Aalto University Department of Radio Science and Engineering, Espoo, Finland\\
$^{34}$ Cahill Center for Astronomy \& Astrophysics, Caltech, 1200 E. California Blvd, Pasadena, CA, 91125, U.S.A.\\
$^{35}$ Department of Physics, Purdue University, 525 Northwestern Ave, West Lafayette, IN 47907, USA\\

\label{lastpage}
\end{document}